\documentclass{blois}

\usepackage{amsmath,amssymb}

\def\be{\begin{equation}}
\def\ee{\end{equation}}
\def\bea{\begin{eqnarray}}
\def\eea{\end{eqnarray}}

\newcommand{\Photo}{}

\begin{document}
\vspace*{4cm}
\title{GRAVITATIONAL PRODUCTION OF MATTER AND RADIATION DURING REHEATING}

\author{ SIMON CLERY}

\address{Universit\'e Paris-Saclay, CNRS/IN2P3, IJCLab, 91405 Orsay, France.}

\maketitle\abstracts{I present the production of matter and radiation during reheating after inflation, considering only gravitational interactions between the inflaton background and the other sectors. Processes considered are the following: i) the exchange of a graviton, $h_{\mu \nu}$, involved in the scattering of the inflaton or particles in the newly created radiation bath; ii) scattering of the inflaton background and particles in the radiation bath including the effects of non-minimal couplings to curvature of the Higgs boson and the inflaton. Requiring the existence of heavy right-handed neutrinos (RHN), I show that a minimal scenario utilizing only these "gravitational portals" is able to generate {\it simultaneously} the observed relic density of Dark Matter (DM), the baryon asymmetry through leptogenesis, as well as a sufficiently hot thermal bath after inflation, for generic models of large field inflation.}

\section{Introduction}

The minimal irreducible interaction that should exist between DM and the Standard Model (SM) is mediated by gravitons which can lead to the observed amount of DM \cite{Mambrini:2021zpp,Clery:2021bwz,Clery:2022wib,Barman:2022qgt}. As this coupling is unavoidable, it provides a lower limit on the number of particles produced. At the end of inflation, during the period of reheating, the available energy in the inflaton field leads to significant gravitational production, which can be more efficient than the production from the thermal bath \cite{Mambrini:2021zpp,Clery:2021bwz}. The radiation bath itself may also be produced during reheating from the inflaton through gravitational effects \cite{Clery:2021bwz,Clery:2022wib,Barman:2022qgt,Co:2022bgh,Haque:2022kez}. However, reheating from graviton exchange processes alone requires a very steep inflaton potential after inflation, resulting in a low reheating temperature and an important enhancement of the primordial power spectrum of tensor modes \cite{Barman:2022qgt,Giovannini:1998bp,Co:2021lkc}. Hence, the minimal scenario of gravitational reheating is excluded by excessive generation of dark radiation in the form of gravitational waves (GWs) at BBN time \cite{Yeh:2022heq}. This limitation of minimal gravitational reheating is one motivation to introduce, as a natural generalization, non-minimal couplings to gravity, especially of the SM Higgs field. This non-minimal coupling enhances the reheating temperature reached through these gravitational portals \cite{Clery:2022wib,Barman:2022qgt,Co:2022bgh}. Besides the necessity of having a DM candidate and inflation, it is well-known that the baryonic content of the Universe is asymmetric. One mechanism to produce the baryon asymmetry of the Universe (BAU) via the lepton sector is called leptogenesis \cite{Fukugita:1986hr}. We show that it is possible to consider a model-independent theory of non-thermal production of RHNs during reheating, from the inflaton background \cite{Barman:2022qgt,Co:2022bgh}. The abundance of RHNs can further lead to observed BAU from the out-of-equilibrium CP violating decay of the RHNs.
In this perspective, we derive a simultaneous solution for the DM abundance, the baryon asymmetry, and the origin of the thermal bath from purely gravitational interactions \cite{Barman:2022qgt}. We further show that the present framework can give rise to a detectable inflationary GWs background for futuristic GWs detectors \cite{Barman:2022qgt}.

\section{The framework}
We study the universal gravitational interactions between the inflaton background, the SM, and the dark sectors. The space-time metric can be expanded locally around flat space using $g_{\mu \nu} \simeq \eta_{\mu \nu} + \frac{2h_{\mu \nu}}{M_P}$ \footnote{$M_P = (8 \pi G_N)^{-1/2} \simeq 2.4 \times 10^{18}$ GeV is the reduced Planck mass.}, where $h_{\mu\nu}$ are canonically normalized gravitons. Gravitational interactions mediated by gravitons are described by the Lagrangian \cite{Clery:2021bwz}
\be
\sqrt{-g}{\cal L}_{\rm min}= -\frac{1}{M_P}h_{\mu \nu}
\left(T^{\mu \nu}_{SM}+T^{\mu \nu}_\phi + T^{\mu \nu}_{N_i} \right) \, .
\label{Eq:lagrangian_min}
\ee
Here SM represents Standard Model fields, and $\phi$ is the inflaton. In the following, we consider $N_i$ to be spin 1/2 Majorana fermions which can be associated with RHNs, accounting for DM or participating in the leptogenesis process. As minimal gravitational production can be insufficient, we also consider the possibility that scalar fields have non-minimal couplings to gravity which generate effective couplings between the different sectors. In order not to spoil inflation and to keep the canonical kinetic term for the inflaton during its oscillating phase, we impose a small field limit which in the end constrains $\xi_\phi \ll 1$ \cite{Clery:2022wib}. Thus, the only relevant non-minimal coupling to gravity considered is the one of the Higgs field, $\xi_h$. It is then convenient to redefine the metric field via the appropriate conformal transformation from the Jordan frame, where gravity is modified by this additional non-minimal coupling, to the Einstein frame, where we recover the usual gravity and additional effective couplings between fields. Expanding the Einstein frame action in powers of $1/M_P^2$, we obtain canonical kinetic terms for the fields and deduce the leading-order interactions induced by the non-minimal coupling \cite{Barman:2022qgt,Co:2022bgh}
\begin{equation}
    \label{lag4point}
    \mathcal{L}_{\rm{non-min.}} \; = \; -\frac{M_{N_i}}{2M_P^2} \xi_{h}\, |h|^2\, \overline{N_i^c}N_i \, -\frac{\xi_h}{M_P}|h|^2\left(2V(\phi) -\frac{1}{2}\eta_{\mu\nu}\partial^\mu\phi\partial^\nu\phi \right).
\end{equation}
where $h$ is the Higgs field in unitary gauge, and $M_{N_i}$, the Majorana mass for the RHN $N_i$. We assume three RHNs, $i = 1,2,3$, where $N_1$ is the lightest one. We assume either that $(y_N)_{1i} = 0$ for all $i$ and that $N_1$ is a stable DM candidate, or consider a metastable DM candidate allowing for $N_1$ to decay into neutrinos that could be observed in experiments such as IceCube \cite{Barman:2022qgt}. The two other RHNs, namely $N_{2,3}$ are assumed to be much heavier and they participate in leptogenesis. Besides the Yukawa couplings, the only couplings considered between the SM, the RHNs, and the inflaton are gravitational of the form in Eq.~(\ref{Eq:lagrangian_min}).\\

For the production of RHNs and radiation through the scattering of the inflaton condensate, we consider the time-dependent oscillation of a classical inflaton field $\phi(t)$ \cite{Garcia:2020wiy}.  The computation depends explicitly on inflaton potential, so without loss of generality, we consider a specific model for inflation called the $\alpha$-attractor T-model \cite{Kallosh:2013hoa}  
\begin{equation}
    V(\phi) \; = \;\lambda M_P^{4}\left|\sqrt{6} \tanh \left(\frac{\phi}{\sqrt{6} M_P}\right)\right|^{k} \, ,
\label{Vatt}
\end{equation}
which can be expanded about the origin 
\footnote{This discussion is general and not limited to T-models of inflation, as long as the inflaton potential can be expanded in a similar way around its minimum.}
\begin{equation}
    \label{Eq:potmin}
    V(\phi)= \lambda \frac{\phi^{k}}{M_{P}^{k-4}}, \quad \phi \ll M_{P} \, .
\end{equation}
In this class of models, inflation occurs at large field values ($\phi \gg M_P$). After the period of exponential expansion, the inflaton begins to oscillate about the minimum where the process of reheating begins. During the oscillations, depending on the parameter $k$, the average equation of state of the inflaton fluid is given by  \cite{book} $ w = \frac{k-2}{k+2}$.

\section{Gravitational portals during reheating}
\subsection{Gravitational reheating}
\begin{figure}[htb!]
\centering
\includegraphics[width=0.32\linewidth]{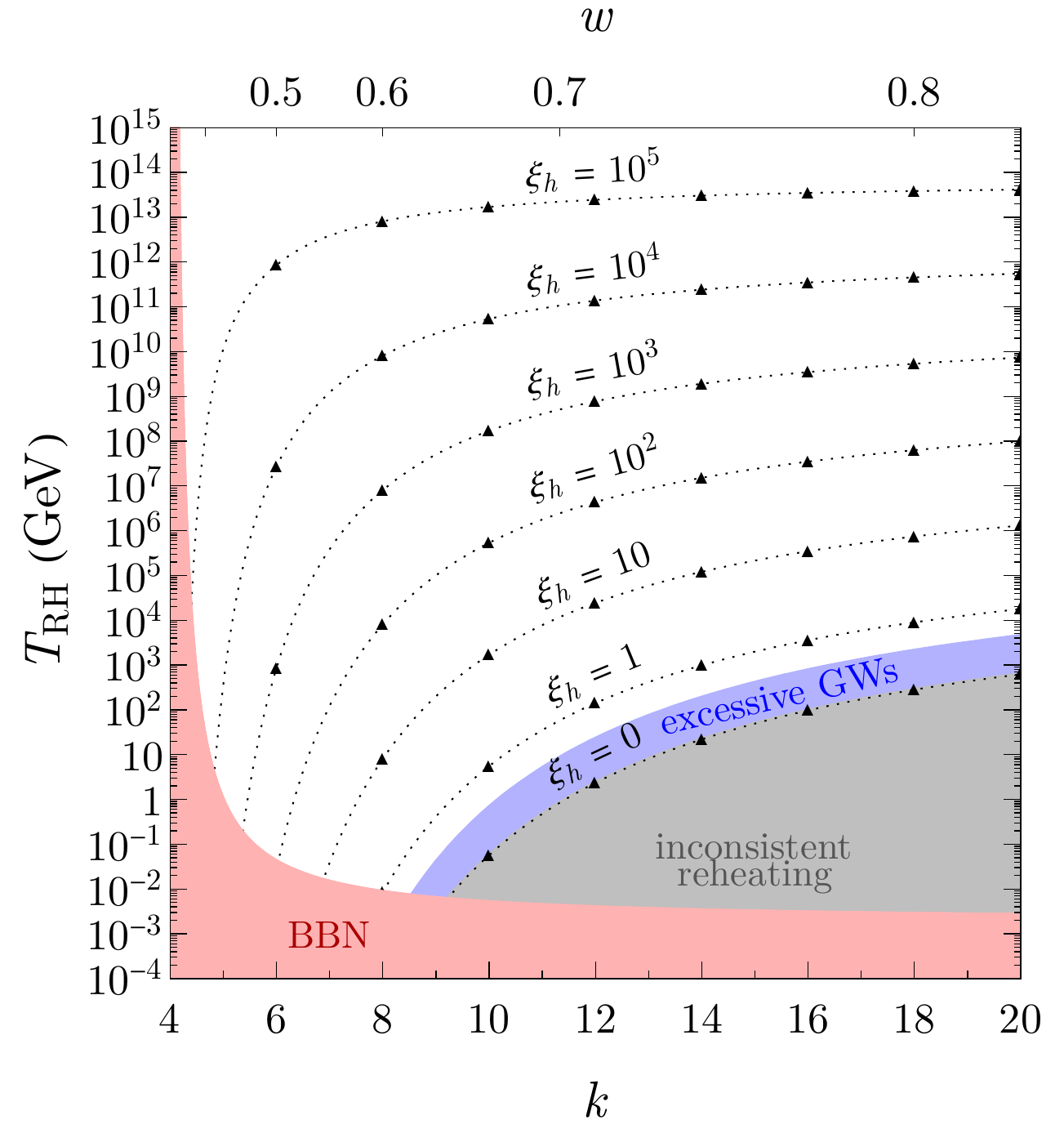}
\includegraphics[width=0.32\linewidth]{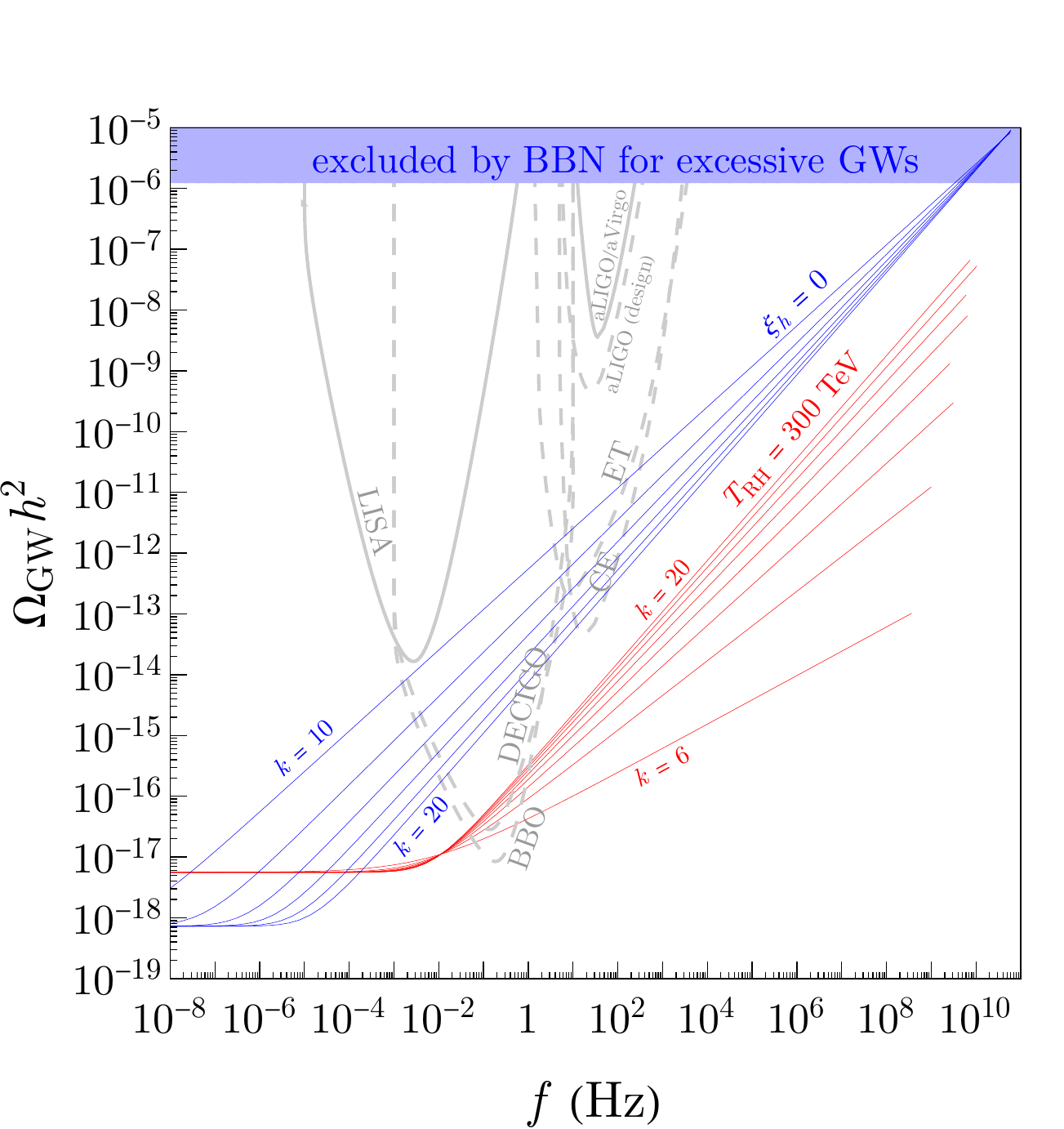}
\includegraphics[width=0.32\linewidth]{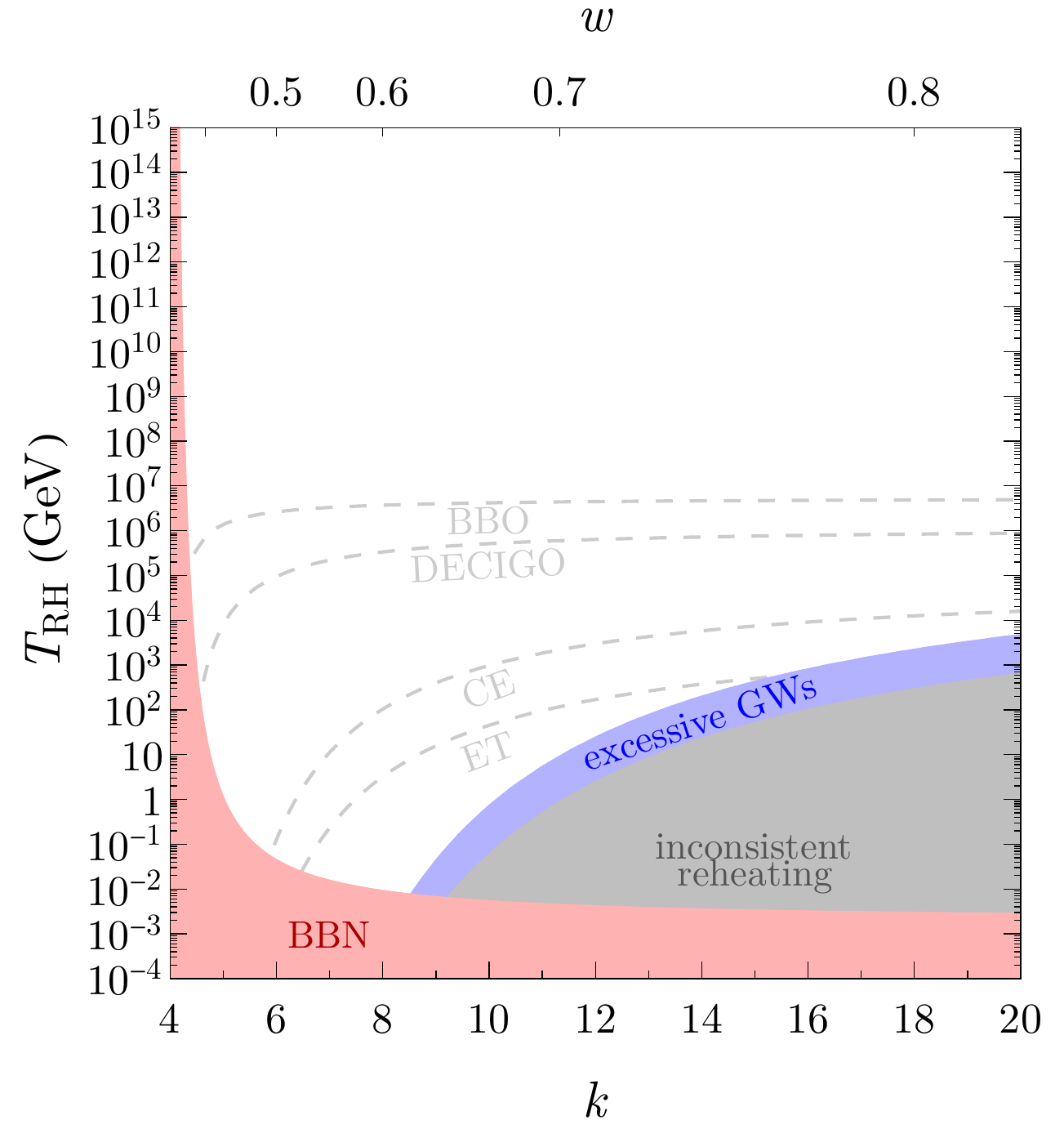}
\caption[]{{\it Left:} Reheating temperature, $T_{\rm RH}$, reached through gravitational production of $h$, including non-minimal contribution from $\xi_h$. The red region is excluded by too low reheating temperature for a successful BBN. {\it Middle:} Relic density of GWs as a function of the frequency, for different ($\xi_h$, $k$) in blue, and fixed $T_{\rm RH}$ with different $k$ in red. The blue region is excluded by BBN for excessive dark radiation. Sensitivity curves of future GWs detectors are shown. {\it Right:} Projection of future GWs detector constraints on the reheating parameter space $(T_{\rm RH}, w)$. The blue region is already excluded by excessive GWs enhancement. \cite{Barman:2022qgt}}
\label{fig:trh}
\end{figure}

We compute the radiation produced in the form of Higgs quanta, during the stage of inflation oscillations. There are two processes that are contributing and interfering: the minimal production coming from graviton exchange between the inflaton and Higgs bosons, and the production due to non-minimal coupling to gravity $\xi_h$ and the interaction described in (\ref{lag4point}). We present in Figure \ref{fig:trh}, the reheating temperature, $T_{\rm RH}$, reached through these gravitational portals as a function of the equation of state parameter $k$, and of the non-minimal coupling, $\xi_h$ \cite{Barman:2022qgt}. The minimal scenario of graviton exchange corresponds to the line $\xi_h=0$. As this process is unavoidable it provides a lower bound on $T_{\rm RH}$, denoted by the grey region of {\it inconsistent reheating}. Besides the production of radiation, an inflation era followed by the reheating where the inflaton redshifts faster than radiation results in an enhancement of primordial GWs generated by quantum fluctuations during inflation \cite{Barman:2022qgt,Giovannini:1998bp,Co:2021lkc}. This enhancement occurs only for tensor modes that crossed the Hubble horizon during the reheating era. Then, for these frequencies, the GW strength is enhanced by a factor of $\rho_\phi / \rho_R$ evaluated at the horizon crossing. As a result, the largest enhancement is for the mode that re-enters the horizon right after inflation. The spectrum exhibits a distinctive enhancement that depends on the shape of inflaton potential near the minimum, $\Omega_{\rm GW}h^2 \propto f^{\frac{k-4}{k-1}}$ \cite{Barman:2022qgt}. This mechanism enhances GWs relic, which contributes to dark radiation. The constraints on the effective relativistic degrees of freedom, $\Delta N_{\rm eff}$, at BBN and CMB decoupling translated into a constraint on GWs relic \cite{Yeh:2022heq} $\Omega_{\rm GW}h^2 \lesssim 5.6\times 10^{-6} \,\Delta N_{\rm eff}$. This constraint can already exclude a part of the parameter space for reheating: a too-low reheating temperature and a too-steep inflaton potential near the minimum (large $k$) lead to an excessive enhancement of GWs spectrum. Thus, non-minimal coupling to gravity is needed to enhance radiation production in this scenario of gravitational reheating. The spectrum enhancement and constraints from BBN \cite{Barman:2022qgt} are presented in the middle panel of Figure \ref{fig:trh}, and the projections of future GWs detector constraints on the reheating parameter space are depicted on the right panel.
\subsection{Gravitational production of DM and Leptogenesis}
\begin{figure}[htb!]
\centering
\includegraphics[width=0.32\linewidth]{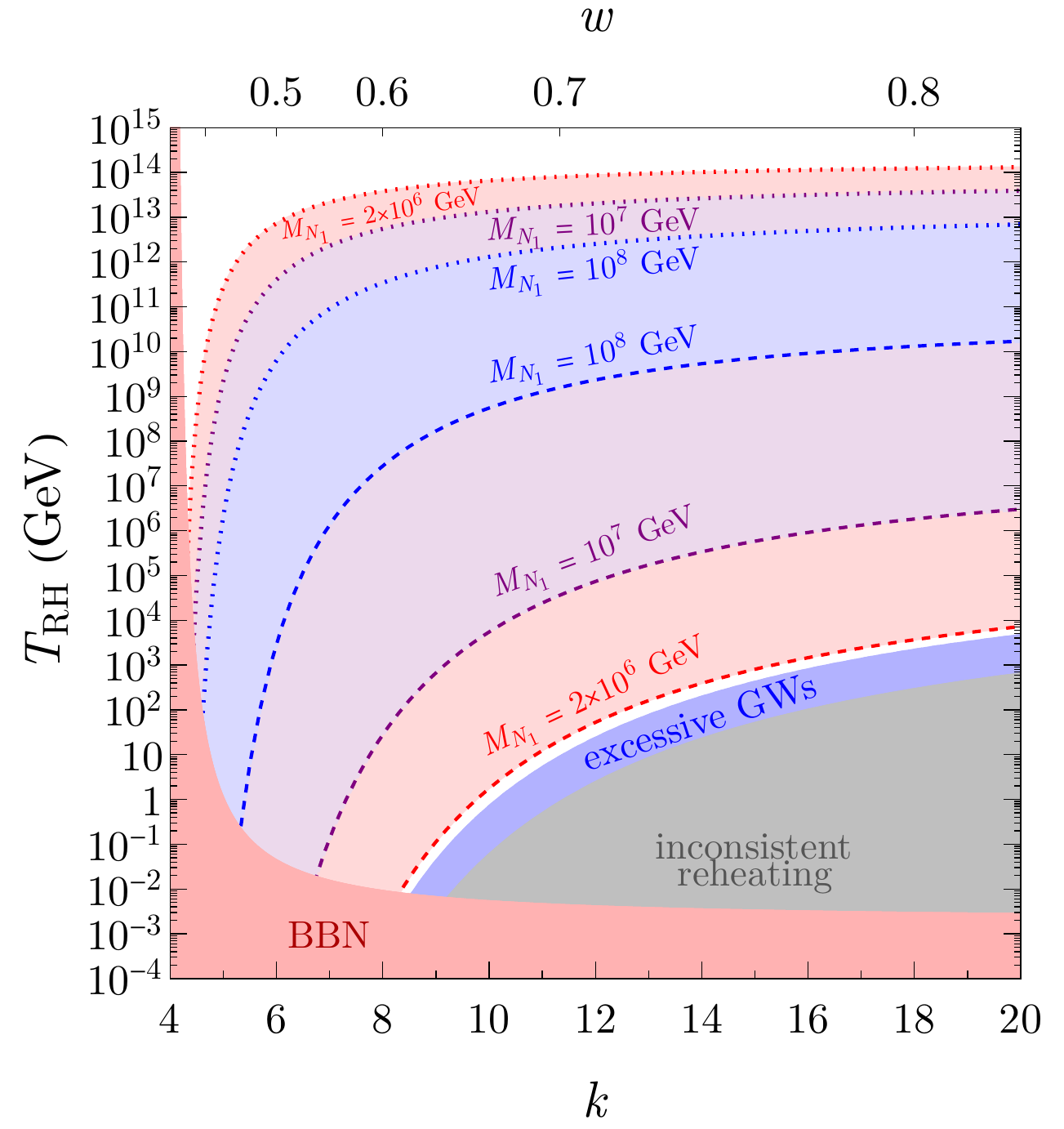}
\includegraphics[width=0.32\linewidth]{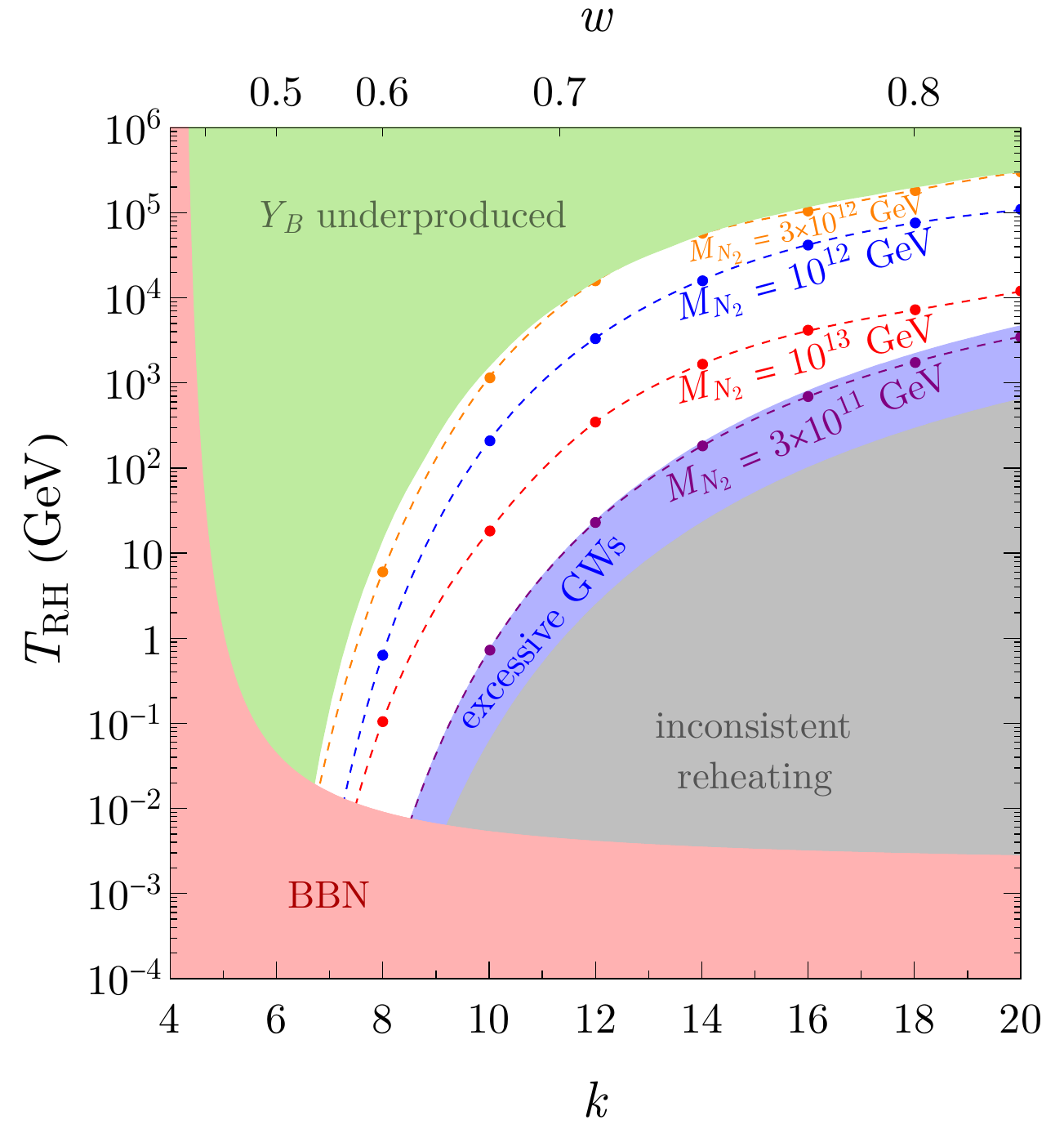}
\includegraphics[width=0.32\linewidth]{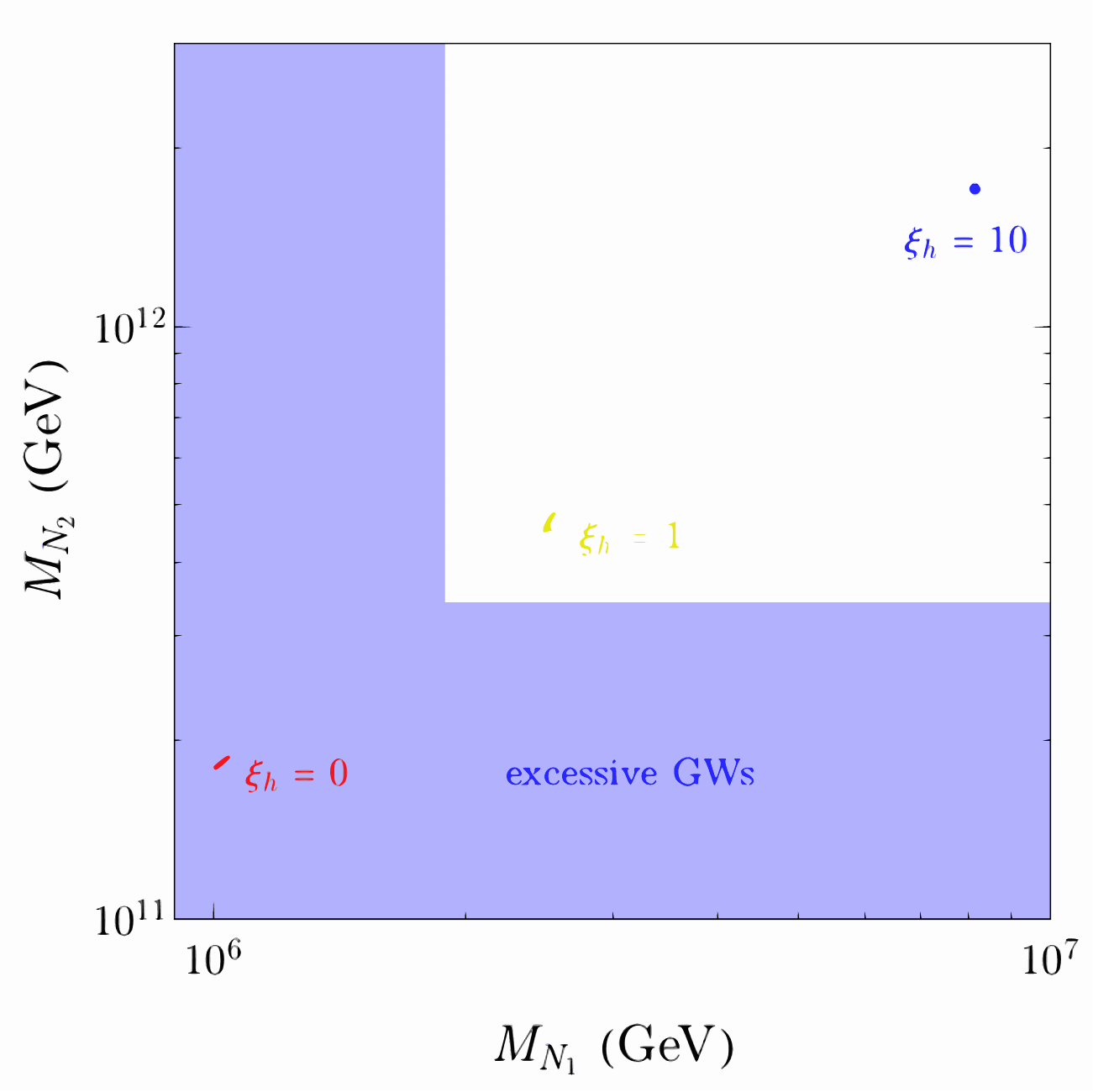}
\caption[]{{\it Left:} Colored regions correspond to values of $(k,T_{\rm RH})$ with $\Omega_{N_1}h^2 \le 0.12$ for three choices of $M_{N_1}$ (red, blue, purple). The line styles indicate the dominant contribution. {\it Middle:} Contours of $M_{N_2}$ corresponding to the observed BAU in the $(k,T_{\rm RH})$ plane. The green region leads to underproduction of $Y_B$ due to the kinematic suppression when $M_{N_2}$ approaches the inflaton mass. {\it Right:} Viable parameter space in the ($M_{N_1}, M_{N_2}$) plane for which gravitational interactions are responsible for the observed DM relic, the BAU (produced from $N_2$ decays), and reheating, for $k \in [6,\,20]$. Different colors correspond to different $\xi_h$. \cite{Barman:2022qgt}}
\label{fig:RHN}
\end{figure}

Similarly, we compute the production of RHNs from gravitational portals, during reheating. 
The production can be sourced either by SM Higgs bosons, exchanging gravitons as well as mediated by the non-minimal coupling to gravity $\xi_h$, or directly from the inflaton background mediated by minimal gravity. 
In Figure \ref{fig:RHN}, we show the constraints on the relic abundances, in the form of a DM component for one of the RHN (left), and participating in the generation of a lepton asymmetry for the two other RHNs (middle). The asymmetry between leptons and anti-leptons originates from the out-of-equilibrium, CP-violating, decays of very heavy RHNs $N_{2,3}$. This lepton asymmetry is subsequently converted into a baryon asymmetry by the $(B+L)$-violating electroweak sphaleron transitions \cite{Kuzmin:1985mm} and can explain the observed BAU $Y_B^{\rm obs} \simeq 8.7\times 10^{-11}$ \cite{Planck:2018vyg}. 
The left plot of Fig. \ref{fig:RHN}, provides regions corresponding to values of $(k,T_{\rm RH})$ with 
$\Omega_{N_1}h^2 \le 0.12$ for three choices of $M_{N_1}$ 
(in different colors). The line styles indicate the 
dominant contribution: thermal (dotted) or from the 
inflaton background (dashed). We see that inflaton 
production of DM dominates for low $T_{\rm RH}$, whereas 
the thermal contribution can dominate for large values of $T_{\rm RH}$ \cite{Barman:2022qgt}. Within each shaded band, the total 
gravitational relic of $N_1$ is below the 
observed DM density and the observed value is reached on the border. As DM mass increases, the viable range in $k$ extends to lower values, and higher reheating temperatures are possible, requiring a larger non-minimal coupling to gravity. In the middle, we show contours for some values of the mass of $N_2$ that explain the observed BAU. We find that the gravitational contribution to the BAU is entirely due to inflaton scattering rather than the thermal contribution, because of the large mass $M_{N_2}$ required \cite{Barman:2022qgt}. Therefore, Leptogenesis via $N_2$ is possible and indicating a mass $M_{N_2} \gtrsim 3 \times 10^{11}$ GeV is required. Larger values of $M_{N_2}$ can produce the correct asymmetry so long as $\xi_h > 0$ to recover sufficiently high gravitational $T_{\rm RH}$. However, when $M_{N_2} \gtrsim 3 \times 10^{12} \rm GeV$, the baryon asymmetry starts to become kinematically suppressed as it reaches the (effective) inflaton mass \cite{Barman:2022qgt}.\\

\subsection{Simultaneous solutions}
Combining the preceding analyses, it is possible for a given $V(\phi)$, to constrain the ($M_{N_1}$, $M_{N_2}$, $\xi_h$) parameter space by requiring leptogenesis, DM production, and reheating to have a common gravitational origin. We project the viable parameter space in the ($M_{N_1}$,$M_{N_2}$) plane in Fig.~\ref{fig:RHN} right pannel, for different values of $\xi_h$, allowing $k$ to vary within $k\in [6,\,20]$. In each colored line segment, gravitational interactions are responsible for the observed DM relic, the BAU and reheating, for a choice of $\xi_h$ \cite{Barman:2022qgt}. The case $\xi_h=0$ is ruled out from overproduction of GWs. 
The maximum possible value for $\xi_h$ is around 13.5, above which the mass $M_{N_2}$ gets too close to the inflaton mass. Interestingly, the viable parameter space is approximately independent of $k$.

\section*{Acknowledgments}

I am grateful to K. Olive, S. Verner, B. Barman, and R. Co for our fruitful collaboration, and very thankful to Y. Mambrini, for his great attention to my work and his extremely wise advice.

\end{document}